\documentclass{article}
\input epsf.sty
\input colordvi.sty

\newcommand{\half}{\frac{1}{2}}
\newcommand{\bright}{\begin{flushright}}
\newcommand{\eright}{\end{flushright}}
\newcommand{\bminip}{\begin{minipage}}
\newcommand{\eminip}{\end{minipage}}
\newcommand{\bcent}{\begin{center}}
\newcommand{\ecent}{\end{center}}
\newcommand{\beq}{\begin{equation}}
\newcommand{\eeq}{\end{equation}}
\newcommand{\beqa}{\begin{eqnarray}}
\newcommand{\eeqa}{\end{eqnarray}}
\newcommand{\barr}{\begin{array}}
\newcommand{\earr}{\end{array}}

\newcommand{\nnb}{\nonumber}
\newcommand{\reflef}{(\ref}
\newcommand{\MP}{M_{\rm P}}

\newcommand{\Lmd}{\Lambda}

\newcommand{\psibar}{\overline{\psi}}

\begin{document}
\baselineskip=0.6cm
\bcent
{\Large\bf Accelerating universe and the time-dependent fine-structure constant
}\footnote{Delivered at Joint Discussion 9 {\it Are the fundamental
constants varying with time?} in XXVII IAU General Assembly, 03-14 Aug
2009, Rio de Janeiro, Brazil, in press in IAU 2009 JD9
conference proceedings, MmSAIt, vol.80, eds.  Paolo  Molaro \& Elisabeth Vangioni.}\\
Yasunori Fujii\\
Advanced Research Institute for Science and Engineering, Waseda
University\\[-.5em]
 Tokyo, 169-8555, Japan, Email: fujii@e07.itscom.net\\
\bminip{14cm}
\mbox{}\\[.8em]
{\large\bf Abstract}
\mbox{}\\[0.4em]
Theoretical background of our proposed relation between the accelerating
universe and the time-variability of the fine-structure constant is discussed,
 based on the scalar-tensor theory, with emphases on the intuitive
 aspects of underlying physical principles.  An important comment is
added on the successful understanding of the size of the effective
cosmological constant responsible for the acceleration, without appealing to fine-tuning parameters.
\eminip
\ecent
\mbox{}\\[-2.4em]

\section{Introduction}

We start with assuming a gravitational scalar field as the dark-energy supposed to be responsible for the accelerating universe.  Also from
 the point of view of unification, a scalar field implies a
time-variability of certain constants observed in Nature.  In this
context we once derived a relation for the time-variability of the
fine-structure constant $\alpha$:
\beq
\frac{\Delta\alpha}{\alpha}=\zeta {\cal Z}\frac{\alpha}{\pi}\Delta\sigma,
\label{prs_3_0}
\eeq
as was detailed in Chapter 6.6 of \cite{cup}, where $\sigma$ is the
scalar field in action in the accelerating universe.  Then we compared the dynamics of the accelerating
universe, on one hand, and $\Delta\alpha/\alpha$ derived from QSO, Oklo
and atomic clocks, on the other hand \cite{yfPL1, yfPL2}.  In this
article we discuss its theoretical background based on the scalar-tensor
theory invented first by Jordan in 1955 \cite{jordan}, focusing upon
the underlying physical principles. For details, see  our references
\cite{cup, ptp}.

\section{Scalar-tensor theory}

The basic Lagrangian is given by
\beq 
{\cal L}\!\!=\!\!\sqrt{-g}\left(\! \half \xi \phi^2 R\! -\!\epsilon \half
\left( \nabla\phi \right)^2 \!\!+\!\!L_{\rm m}\! -\!\Lmd\! \right),
\label{prs_3}
\eeq 
where $\phi$ is the scalar field, while $(\nabla\phi )^2\!
=g^{\mu\nu}\partial_\mu\phi \partial_\nu\phi$.  The two parameters  
$\epsilon, \xi$ are related to the better known symbol $\omega$ by  
$\epsilon ={\rm Sgn (\omega)}, 4\xi =|\omega|^{-1}$.  We also included
$-\Lmd$ as the simplest extension from the original scalar-tensor theory.
The first term is a well-known nonminimal coupling term 
defining an  effective gravitational constant $8\pi G_{\rm eff}=(\xi
\phi^2)^{-1}$.

$L_{\rm m}$ is the matter Lagrangian for which  Brans  and  Dicke
\cite{bd} added an assumption that the field $\phi$ never enters  $L_{\rm
m}$, because they could save the idea of Weak Equivalence Principle
(WEP) only in this way. Since then, their proposal combined with Jordan's original theory has
been known widely as the Brans-Dicke theory.  We prefer, however, to use a
more modest name like the BD model, partly because it seems likely to be
replaced ultimately by another model for a better understanding of the
accelerating universe, as will be argued later in this article.

Also for the later convenience, let us give an example of $L_{\rm m}$ for
a free massive Dirac field as a convenient representative of matter fields; 
\beq
 L_{\rm m} =-\psibar \left( \partial\hspace{-.5em}/  +m\right)\psi. 
\label{prs_7}
\eeq 
Due to the assumed absence of $\phi$, the mass of this Dirac  particle
 is simply $m$, a pure  constant.  By extending this  simple  argument
 we find a general (and traditional) rule that the  constancy of masses
 of matter  particles is a unique  feature of the BD  model.

We use the reduced Planckian units defined by $c=\hbar =\MP(=
(8\pi G)^{-1/2})=1$.  As an example, the present age of the universe $t_0 \approx 1.4\times 10^{10} {\rm y}$ can be re-expressed as $\sim 10^{60}$.

We apply what is known as the conformal transformation defined by $
g_{\mu\nu}\rightarrow  g_{*\mu\nu} =\Omega^2 g_{\mu\nu}$, where $\Omega(x)$ is an arbitrary spacetime function.  This allows us 
 to  re-express the Lagrangian \reflef{prs_3}) now in terms of the new
 transformed metric  $g_{*\mu\nu}$.  A special choice $\Omega^2  =\xi
 \phi^2$ leads to  the particularly simple  result;

\beq
{\cal L}\!=\!\!\sqrt{-g_{*}}\left(\!\half R_{*}\!\! -\!
\half\left(\nabla_*\sigma\right)^2\!\!-\! \!V(\!\sigma \!)\!+\!L_{\rm *{\rm m}}\!  \right),
\label{prs_5}
\eeq
where the $R_*$ term is  multiplied by a  pure constant, no nonminimal
coupling term, hence the  same as in the Einstein-Hilbert term.  For
this reason we say we have moved to the Einstein (conformal) frame. We also 
say that we come from the  Jordan frame, so-called widely.  We learn a
lesson that the effective gravitational constant can be constant or
variable depending on what frame is chosen.

The scalar field $\phi$ in \reflef{prs_3}) has been replaced in
\reflef{prs_5}) by $\sigma$ which are related to each other by $\phi
=\xi^{-1/2}e^{\zeta\sigma}$ where $\zeta =(6+\epsilon\xi^{-1})^{-1/2}$.  
Notice, unlike $\epsilon$ in \reflef{prs_3}), positivity of the
 kinetic-energy term of the {\em diagonalized} $\sigma$ is assured by
$\zeta^2 >0$ even with a negative $\epsilon$ \cite{cup, ptp}.

Also a constant term $\Lmd$ in \reflef{prs_3}) has been converted to a
potential $V(\sigma) =\Lmd e^{-4\zeta\sigma}$.  Otherwise we put the
asterisks almost everywhere.  Again for the later convenience  in
discussing cosmology, we add 
\beq
a_*=\Omega a, \quad \mbox{and}\quad dt_*=\Omega  dt. 
\label{prs_5_1}
\eeq
According to the first relation, the way of the cosmological expansion
may differ from frame to frame.  The second equation expresses a
coordinate transformation of $t$ to the cosmic time $t_*$ re-defined in the
Einstein frame.

The matter fields are also transformed according to
\beq
\psi_* =\Omega^{-3/2}\psi, \quad \mbox{with}\quad  m_*=\Omega^{-1} m,
\label{prs_5_2}
\eeq
which yields
\beq
 L_{*\rm m} =-\psibar_* \left( \nabla_*\hspace{-1.1em}\mbox{\raisebox{.2em}{$/$}} \hspace{.4em} +m_*\right)\psi_*, 
\label{prs_5_3}
\eeq
showing a form-invariance as compared with \reflef{prs_7}), though we
ignored some complications as discussed in Appendix F of \cite{cup}.

We may compare \reflef{prs_3}) with the effective
Lagrangian for the closed strings formulated in higher-dimensional
spacetime as shown by  Eq. (3.4.58) of \cite{gsw};
\beq
{\cal L}_{\rm st}\!=\!\!\sqrt{-g}e^{-2\Phi}\left( \half R+2  \left( \nabla \Phi\right)^2
+\cdots \right), 
\label{prs_st}
\eeq
with the unique occurrence of a scalar field $\Phi$ called dilaton.
By introducing $\phi$ by $\phi =2e^{-\Phi}$, we
re-express this effective Lagrangian, part of which agrees
 precisely with the first two terms of \reflef{prs_3}) with the
 choice; $\epsilon =-1, \xi=1/4$, or $\omega=-1$.   In this sense we
 might  call the Jordan frame as the string frame or the theoretical
 frame, suggesting that the Jordan frame represents a world in which
 unification is realized.  This might provide a rationale to introduce
 a constant $\Lmd$ of the Planckian size in \reflef{prs_3}).   Then how
 about the observational or physical frame?  According  to Dicke
 \cite{dicke} in this connection, the  conformal transformation is a
 local  change of units. Let us emphasize this view.

Suppose we use an atomic clock, measuring time in reference to a
frequency of certain atomic transition, in which we have the
fundamental unit based on $m_{\rm e}$, the electron mass, ignoring the
realistic choice of the reduced mass, for simplicity.  Then we find we
have no way to 
detect any change, if any, of $m_{\rm e}$ itself,  as far as we 
continue to use the atomic clock.  This situation might be put in a more
general term as Own Unit Insensitivity Principe (OUIP) \cite{ptp}. We
then come to 
say that using an atomic clock implies that we are in a physical frame
in which $m_{\rm e}$ is kept constant. According to what we discussed
about the BD model particularly in connection with the constancy of the masses,
we find ourselves in the Jordan frame, which is then identified with the
physical frame.

The above simple argument can be extended to 
astronomical observations based on measuring redshifts of atomic
spectra, in which we also use $m_{\rm e}$ as a fundamental unit in the
same way as in using an atomic clock.  In this context, we repeat the
same statement on the physical frame as the Jordan frame, as far as we accept the BD model.  But
the situation will be subject to change in the presence of $\Lmd$.

\section{Cosmology}

We now discuss cosmology in the presence of $\Lmd$ assumed to be
positive with the Planckian size, first in the Jordan frame.  Under usual
simplifying assumptions on the metric  also in radiation-dominance, for
the moment, we may assume the spatially uniform $\phi$ depending only on the
cosmic time $t$.  We then write down the cosmological equations,
obtaining asymptotic solutions, 
\beqa
a&=& \mbox{const}, \nnb\\
\phi &=& \sqrt{\frac{4\Lmd}{6\xi +\epsilon}}\:t, \nnb \\
\rho &=&-3\Lmd \frac{2\xi+\epsilon}{6\xi +\epsilon}, 
\label{prs_solnJ}
\eeqa
where $a$ is the scale factor while $\rho$ is the matter density.

Most striking is the first line describing a {\em static} universe.  We
emphasize that \reflef{prs_solnJ}) is a set of the attractor solutions,
which any solutions starting with whatever initial values tend to, as
confirmed by our recent reanalysis \cite{kmyf}.  In this sense we  can
hardly accept the  Jordan frame as a realistic physical frame, {\em contrary}
to what we  concluded toward the end of the preceding section.

We also add that our solution shows no smooth behavior in the limit
$\Lmd \rightarrow 0$.  In other words, in the presence of $\Lmd$ the
solution could be quite different from what had been known in its
absence.  The presence of $\Lmd$ may result in a drastic change.  So how
about in the Einstein frame?

The solution can be obtained either by  starting from the cosmological
equations derived from \reflef{prs_5}), or by applying the conformal
transformation directly to those solutions given by \reflef{prs_solnJ}).
The result is
\beqa
a_* &=& t_*^{1/2}, \nnb\\
\sigma &=& \bar{\sigma}+\half \zeta^{-1}\ln t_*, \nnb \\
\rho_\sigma &=& \frac{3}{16}t_*^{-2}, \nnb \\
\rho_* &=&\frac{3}{4}\left(1-\frac{1}{4}\zeta^{-2}  \right)t_*^{-2},
\label{prs_solnE}
\eeqa
where $\bar{\sigma}$ is given by $\Lmd e^{-4\zeta\bar{\sigma}}=(16
\zeta^2)^{-1}$, while the energy densities $\rho_\sigma$ and $\rho_*$ are
for $\sigma$ and the matter, respectively. The first line, due to the
relations in \reflef{prs_5_1}), shows that the universe does expand,
precisely in the same way as in the ordinary radiation-dominated
universe.  We might be tempted to accept the Einstein frame as the
physical frame.  By comparing the first of \reflef{prs_5_1}) and the second
of \reflef{prs_5_2}), however, we find
\beq
m_*\sim t_*^{-1/2}.
\label{prs_solnE_m}
\eeq
This is not constant in contradiction with OUIP, which would have been
respected only if $m_*$  were to be constant.  The universe looks again
unrealistic;  we have no way to accommodate a physical frame which
should be realistic.

We trace the  origin of this difficulty back to the result of the
static universe in the Jordan frame combined with constant mass of
matter fields as discussed immediately after \reflef{prs_7}) thus
yielding $am =a/m^{-1} =$ const, implying that the universe expands in
the same rate as the microscopic length scale provided by the particle
mass.  This is also likely carried over to the Einstein frame; $a_*m_*
=$ const, which is totally {\em inconsistent} with the current view on today's
cosmology.

We naturally wondered if we can find a way out by somehow forcing $m_*$ to stay
constant, still maintaining the expansion of $a_* \sim t_*^{1/2}$ as it
is so that we could accept the Einstein frame as a physical frame.  We
finally decided to {\em leave} the BD model.  We revised  our previous way to
determine $L_{\rm m}$, replacing the mass term in \reflef{prs_7}) by the
Yukawa-coupling term $-f\psibar \psi\phi$.  Then we do find a constant
$m_*$, generated spontaneously, hence achieving our expected goal, the
Einstein frame identified 
with the physical frame.  The coupling constant $f$ is dimensionless.  This
simple  feature is  shared by any other terms in the basic Lagrangian
\reflef{prs_3}), except for the $\Lmd$ term.  In this sense we have a
global scale invariance, though partially, so the name, the
{\em scale-invariant} model to {\em replace} the BD model.

As a price, however,  we allowed the $\phi$ to enter $L_{\rm m}$
against what BD assumed.  So we should expect WEP violating terms, which
turn out fortunately unobservable in the classical  limit according to
the scale-invariant model; constant masses imply their decoupling from
$\sigma$ depriving matter particles of their role to detect the
violating effects.  But there are quantum  effects as well
arising from the interactions among matter fields,  regenerating the
violating terms, which are found to be somewhat suppressed  according to the
estimate by means of quantum anomalies, a  well-established technique in
the  relativistic quantum field theory.  This is precisely the way the
relation \reflef{prs_3_0}) has been derived.

 We point out  that the  crucially important ingredients were the simple
 and straightforward  arguments on how to define a physical conformal
 frame in the  presence of $\Lmd$.   Also to be re-iterated is that  the
 entire result  hinges upon the  static universe encountered in the
 attractor solution for the radiation-dominated universe in the Jordan
 frame, though  this rather unexpected behavior is found to be somewhat
 exceptional.  In fact we face even more complicated situation for the
 dust-dominated universe, but  with the same final  result in terms of the
 scale-invariant model.   For  more details, as well  as for the ensuing
 phenomenological analyses,  see our references  \cite{cup, ptp, yfPL1,
 yfPL2}.

This scale-invariant model turns out to show an advantage in the present
approach.  It allows an interpretaion of $\sigma$ as a Nambu-Goldstone
boson of dilatation symmetry, massless dilaton, to be consistent with
the absence of the scalar mass term in the starting Lagrangain \reflef{prs_3}),
assumed widely but rarely on any ground stated explicitly.  Quantum
effects mentioned above eventually break the symmetry reducing $\sigma$
to a pseudo-NG boson, still making it easy to understand why WEP
violation is supposed to be mediated by a scalar field, likely as light
as $\sim 10^{-9}{\rm eV}$, or the force-range of macroscopic distances
$\sim 100{\rm m}$, up to a latitude of a few orders of magnitude, as
suggested before \cite{intmed}, revisited again in Chapter 6.4 of
\cite{cup}.  Also in this context, the same quantum anomaly effect 
is responsible for $\Delta\alpha/\alpha$, which by itself may not look
relevant directly to the WEP violation, whereas the same type of
technique can be applied to derive the time-dependent mass ratio $\mu$
between proton and electron, which is obviously WEP violating as far as
$\sigma$ participates in gravitational phenomena.

\section{A comment}

Finally we emphasize that the present approach is related closely to the
question how successfully we understand the accelerating universe.  Of
 central importance is, as was discussed in \cite{succ}, the size of
$\Lmd_{\rm eff}$ required to fit 
the observed acceleration, as small as, $\sim 10^{-120}$ in the
Planckian units, a well-known number which has symbolized what is
known as a fine-tuning problem, a manifestation of  our uneasiness if
our theory is good enough to the  accuracy of as much as 120
orders of magnitudes.  But a numerical similarity
\beq
\Lmd_{\rm eff} \sim t_{*0}^{-2},
\label{prs_50}
\eeq
with $t_{*0}\sim 10^{10}{\rm y}\sim 10^{60}$ appears too remarkable to
be dismissed as a mere coincidence.  Note that this result had been
foreseen in \cite{yf82, bert} though based on somewhat different
interpretation of the scalar-tensor theory including $\Lmd$.

In fact our cosmological solution in the Einstein frame contains the
third line of \reflef{prs_solnE});
\beq
\rho_\sigma \sim t_{*}^{-2},
\label{prs_51}
\eeq
where LHS is the density of the dark-energy, and can be interpreted as
$\Lmd_{\rm eff}$.   This might be called Scenario of a Decaying
Cosmological Constant, providing with an immediate justification of the
relation \reflef{prs_50}).  To be noticed is that today's value of LHS
of \reflef{prs_51}) is small nearly automatically simply because we are old 
enough  as indicated on RHS, requiring no need for an extreme and
unnatural fine-tuning  processes, somewhat reminiscent of Dirac's argument \cite{dirac}. This simplicity and naturalness 
deserves to be called a major success of the scalar-tensor theory in its
simplest extension by introducing $\Lmd$ in the Jordan frame, unparalleled by any other phenomenological approaches.

But we concede that this argument applies only to the  overall  behaviors of
 the universe.   Another kind of  mechanism has to be worked out  for
 non-overall  behaviors, including  the extra acceleration of the
 universe as we  see it.  We may have a choice, but in such a way to
 inherit what we  called the  success.  The present author expects an
 oscillation of $\sigma$ \cite{cup, yfPL1, yfPL2} then the same  in the
 observed  $\Delta\alpha/\alpha$ and  $\Delta\mu/\mu$  to a better  precision
 hopefully to be  realized in  the near future.

\bibliographystyle{aa}

\end{document}